\documentclass[aps,prl,floatfix,twocolumn]{revtex4}
\usepackage{graphics}
\usepackage{dcolumn}
\usepackage{epsfig}

\begin{document}

\title
{Energy Gaps and Stark Effect in Boron Nitride Nanoribbons}
\author{Cheol-Hwan Park}
\author{Steven G. Louie}
\email{sglouie@berkeley.edu}
\affiliation{Department of Physics, University of California at Berkeley,
Berkeley, California 94720\\
Materials Sciences Division, Lawrence Berkeley National
Laboratory, Berkeley, California 94720}

\date{\today}

\begin{abstract}
A first-principles investigation of the electronic properties of
boron nitride nanoribbons (BNNRs)
having either armchair or zigzag shaped edges
passivated by hydrogen with widths up to 10 nm
is presented.
Band gaps of armchair BNNRs exhibit family-dependent oscillations
as the width increases and, for ribbons wider than 3~nm,
converge to a constant value that is 0.02~eV smaller
than the bulk band gap of a boron nitride sheet
owing to the existence of very weak edge states.
The band gap of zigzag BNNRs
monotonically decreases and converges to a gap that is 0.7~eV
smaller than the bulk gap
due to the presence of
strong edge states.
When a transverse electric field is applied,
the band gaps of armchair BNNRs decrease monotonically
with the field strength.
For the zigzag BNNRs, however, the band gaps and the carrier effective
masses either increase or decrease depending on
the direction and the strength of the field.
\end{abstract}
\maketitle

Two-dimensional crystals, including
graphene and single layer of hexagonal boron nitride (BN),
have recently been fabricated~\cite{novoselov:2005PNAS_2D}.
Among them, only graphene has been
studied extensively~\cite{geim:2007NatMat_Graphene_Review}.
Unlike graphene, a hexagonal BN sheet is
a wide gap insulator like bulk hexagonal BN~\cite{blase:1995PRB_hBN_GW}
and is a promising material in optics and
opto-electronics~\cite{watanabe:2004NatMat_hBN}.

\begin{figure}
\includegraphics[width=1.0\columnwidth]{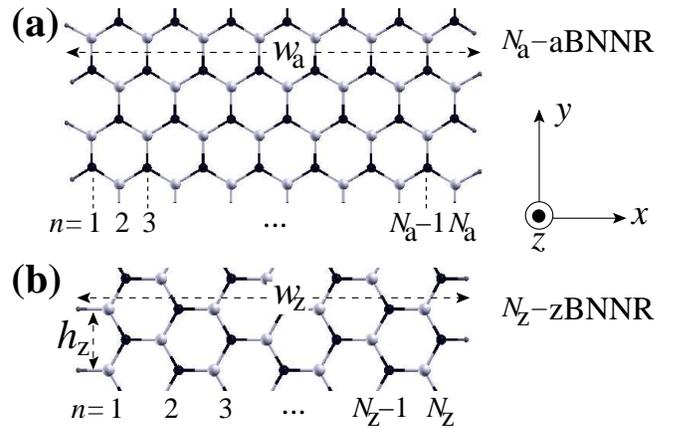}
\caption{Schematic of (a) 14-aBNNR and (b) 7-zBNNR passivated by hydrogen atoms.
Boron, nitrogen and hydrogen atoms are represented by white, black and
grey spheres, respectively. BNNRs are periodic along the $y$ direction.}
\label{Figure_structure}
\end{figure}

Graphene nanoribbons (GNRs)~\cite{nakada:1996PRB_GNR}
with width a few to a hundred nanometers
have been produced by lithographical
patterning~\cite{han:2007PRL_GNR_bandgap,chen:2007PhysicaE_GNR_bandgap}
or chemical processing~\cite{li:2008Science_GNR_bandgap} of graphene.
We expect that boron nitride nanoribbons (BNNRs)
could also be made using similar or other techniques.
Figures~\ref{Figure_structure}(a) and \ref{Figure_structure}(b)
show the structures of
an armchair BNNR with $N_{\rm a}$ dimer lines ($N_{\rm a}$-aBNNR)
and a zigzag BNNR with $N_{\rm z}$ zigzag chains ($N_{\rm z}$-zBNNR),
respectively.
A tight-binding study of the bandstructures of 21-aBNNR
and 13-zBNNR (corresponding to widths $\sim3$~nm)~\cite{chen:2002SSC_BNNR_Opt}
and first-principles investigations of the electronic properties of
small width BNNRs~\cite{nakamura:2005PRB_BNCR,du:2007CPL_BNNR} have been reported.
However, to our knowledge, first-principles calculations on the electronic properties of
experimentally realizable size of BNNRs have not been performed.

Under a transverse electric field,
carbon nanotubes with impurity atoms are expected to show novel band
gap opening behaviors~\cite{son:216602}, whereas zigzag GNRs reveal
half-metallicity~\cite{son:2006GNR_Halfmetal}.
On the other hand, single-walled boron nitride nanotubes (SW-BNNTs),
which are rolled up BN sheets~\cite{rubio:1994PRB_BNNT,
blase:1994EPL_BNNT,chopra:1995Sci_BNNT}, have been predicted to show gigantic
Stark effect in their band gaps in response to a transverse electric field~\cite{khoo:2004PRB_BNNT_Stark},
and this effect has been confirmed experimentally~\cite{ishigami:2005PRL_BNNT_Stark}.
The effect becomes stronger in larger diameter SW-BNNTs~\cite{khoo:2004PRB_BNNT_Stark}.
A similar phenomenon is expected in BNNRs.
Unlike SW-BNNTs, however, BNNRs can be arbitrarily wide.
Therefore, the consequences of the Stark effect in BNNRs would be even more dramatic
than in SW-BNNTs.

In this study, we report first-principles calculations on the electronic
properties of armchair and zigzag BNNRs up to width of 10~nm
with hydrogen passivation of the edge carbon atoms.
We show that the band gaps of armchair and zigzag BNNRs do not
converge to the same value even when
the ribbons are very wide.
The band gap of armchair BNNRs, obtained by density functional theory (DFT)
calculations within the local density approximation (LDA), converges to
a value that is 0.02~eV smaller than the LDA band gap of 4.53~eV of a BN sheet~\cite{note:bulk}.
Unlike armchair GNRs,
the lowest unoccupied band of the armchair BNNRs
is composed of edge states with energy position asymptote to a fixed value when
the ribbon is wider than 3~nm, the decay length of the edge-state.
The band gap of the zigzag BNNRs, also
determined by edge states, is monotonically reduced as a function of increasing width
and converges to a value that is 0.7~eV smaller than the LDA bulk gap
because, as discussed below, of an additional edge polarization charge effect.
The DFT Kohn-Sham eigenvalues within LDA in general underestimate the band gaps
of materials; an accurate first-principles calculation of band gaps
requires a quasiparticle approach~\cite{hybertsen:1986PRB_GW}.
The basic physics discovered here however should not be changed.

When a transverse electric field is applied, the highest occupied and the lowest
empty states in armchair BNNRs become localized at the two different edges.
Because of the external electrostatic
potential difference between the two edges,
the band gap is reduced with increased field strength. On the contrary,
in zigzag BNNRs, depending on the field direction,
the states near the band gap either become more localized at the edges
or less so. Also, the band gaps and effective masses either decrease or increase
depending on field strength and direction.
These novel properties could be used in manipulating the transport
properties of doped BNNRs.

We performed {\it ab initio} pseudopotential DFT calculations
within LDA in a supercell configuration using the SIESTA
computer code~\cite{portal:1997siesta}.
A double-zeta plus polarization basis set was used and ghost orbitals
were included to describe free-electron-like
states~\cite{blase:1995PRB_hBN_GW,khoo:2004PRB_BNNT_Stark,park:2006PRL_BNNT_Opt}.
A charge density cutoff
of 300~Ry was used and atomic positions were relaxed so that the force on
each atom is less than 0.04~eV/\AA. To eliminate spurious interactions
between periodic images, a supercell size
of up to  20~nm$\times$20~nm in the $xz$ plane was used.

\begin{figure}
\includegraphics[width=1.0\columnwidth]{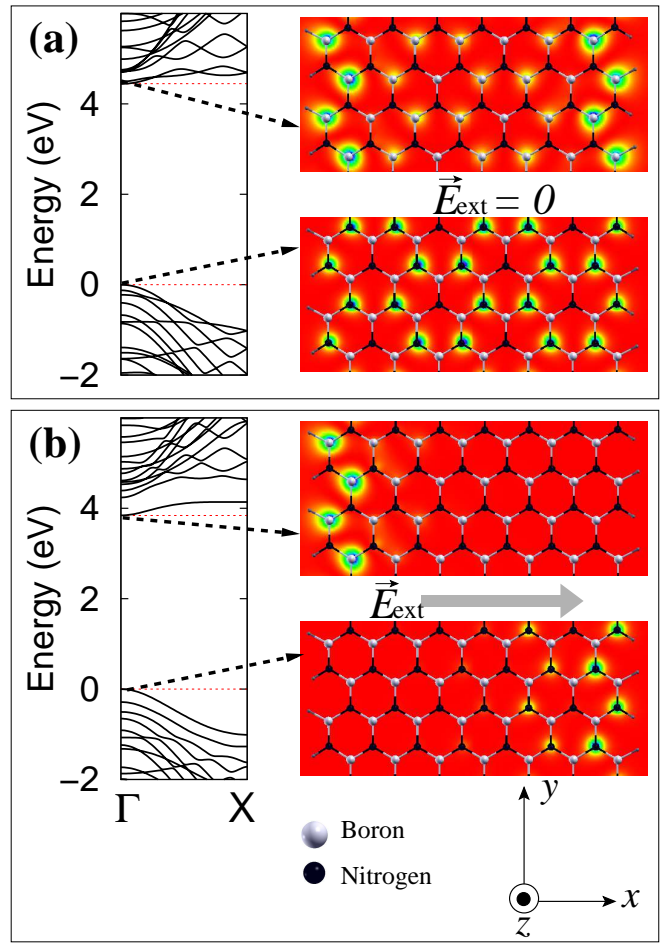}
\caption{(color online) LDA energy bandstructure (left) and the squared wavefunctions
integrated along $z$ of the highest occupied
state (right lower) and the lowest unoccupied state (right upper) of 14-aBNNR
under an external electric field $\vec{E}_{\rm ext}$ of strength
(a) zero and (b) 0.1 eV/\AA\ along $+x$ direction.
Dashed red lines in the bandstructure indicate
the energies of the band edge states.
In the wavefunction plots, green regions are
associated with high densities.}
\label{Figure_a14BN}
\end{figure}

The armchair BNNRs are found to have a direct gap
at the zone center [left panel of Fig.~\ref{Figure_a14BN}(a)].
The highest occupied state, the valence band maximum (VBM),
has a wavefunction which
is localized at nitrogen atoms throughout the ribbon
[right lower panel of Fig.~\ref{Figure_a14BN}(a)].
The lowest empty state, the conduction band minimum (CBM),
is however an edge state with wavefunction localizing
at the boron atoms on the edges
[right upper panel of Fig.~\ref{Figure_a14BN}(a)]. In contrast,
the corresponding state for an armchair GNRs is delocalized throughout the
ribbon~\cite{brey:2006PRB_GNR}.
The total potential near the edge of the armchair BNNRs is different from that of
the bulk~\cite{park:2007UNP_BNNR}.
By incorporating this variation into the {\it on-site potential energies}
of a few BN dimer lines near the edges,
one could reproduce the main features of the states near the band gap
within a tight-binding formulation~\cite{park:2007UNP_BNNR}.

\begin{figure}
\includegraphics[width=0.95\columnwidth]{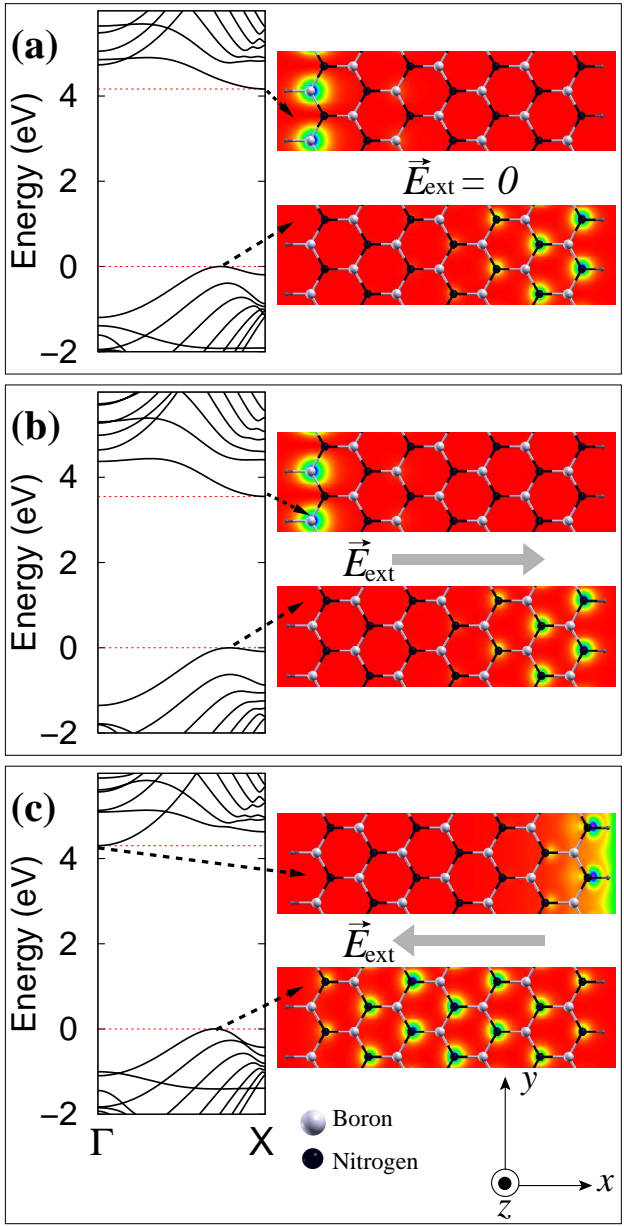}
\caption{(color online) LDA energy bandstructure (left) and the squared wavefunctions
integrated along $z$ of the highest occupied
state (right lower) and the lowest unoccupied state (right upper) of 7-zBNNR
under an external electric field $\vec{E}_{\rm ext}$ of strength
(a) zero, (b) 0.1 eV/\AA\ and (c) $-0.1$ eV/\AA\ along the $x$ direction.
Dashed red lines in the bandstructure indicate
the energies of the band edge states.
In the wavefunction plots, green regions are
associated with high densities.}
\label{Figure_z7BN}
\end{figure}

The zigzag BNNRs have the VBM at a point between the X and the $\Gamma$ points
which has wavefunction localized at the nitrogen edge and the CBM at the X point
which has wavefunction localized at the boron edge [Fig.~\ref{Figure_z7BN}(a)],
in agreement with results by Nakamura {\it et al.}~\cite{nakamura:2005PRB_BNCR}

\begin{figure}
\includegraphics[width=1.0\columnwidth]{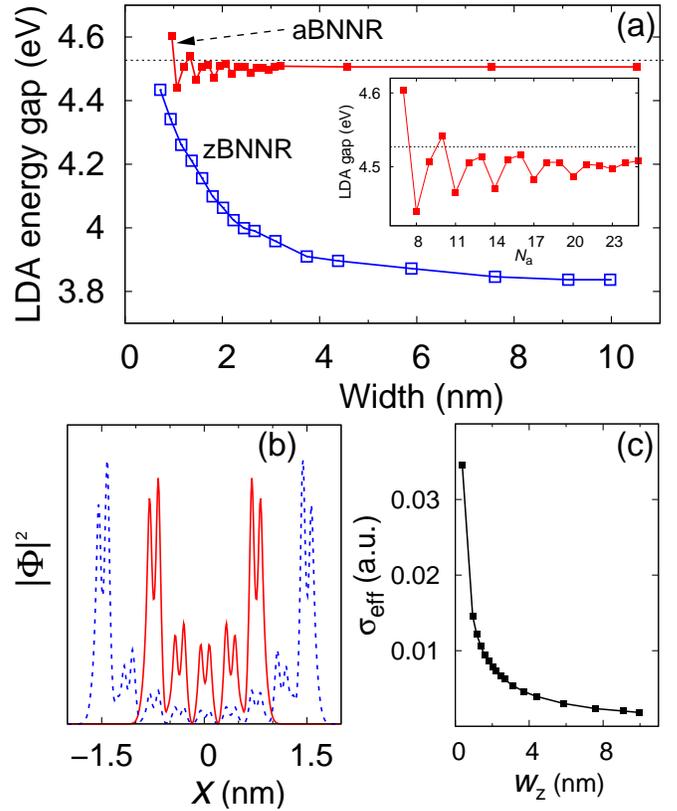}
\caption{(color online) (a) Band gaps of armchair (filled red squares) and zigzag
 (empty blue squares) BNNRs versus their widths.
 Inset: band gaps of armchair BNNRs versus $N_{\rm a}$ (see Fig.~\ref{Figure_structure}).
 Solid lines are a guide to the eyes.
 Dashed lines indicate the bulk band gap of a BN sheet with no edges.
(b) Probability distributions $|\Phi({\bf r})|^2$
integrated in the $yz$ plane (see Fig.~\ref{Figure_structure}) versus
the distance along the $x$ direction from the ribbon center
for the lowest unoccupied
state in 14-aBNNR (solid red line) and 26-aBNNR (dashed blue line).
(c) The effective polarized line charge density $\sigma_{\rm eff}$
of zigzag BNNRs versus $w_z$.
The solid line is a guide to the eyes.}
\label{Figure_bandgap}
\end{figure}

Figure~\ref{Figure_bandgap}(a) shows the band gap variation of the armchair
and the zigzag BNNRs with width.
The most interesting and somewhat counter-intuitive feature is that
as the width increases,
band gaps of the armchair and the zigzag BNNRs converge to different values
neither of which is that of a BN sheet.
This is because the CBM or the VBM
are determined by edge-states.

The edge-state band gaps of armchair BNNRs show a family behavior
with respect to the number of dimer lines $N_{\rm a}$ [inset of Fig.~\ref{Figure_bandgap}(a)],
the family with $N_{\rm a}=3n-1$ having the smallest gaps where $n$ is an integer.
A similar trend is observed in armchair
GNRs~\cite{fujita:1996JPSJ_GNR,brey:2006PRB_GNR,son:2006GNR_LDA,barone:2006GNR,white:2007GNR}.
The edge-state band gaps of the armchair BNNRs
converge to a near constant value roughly when the ribbon is wider
than 3~nm [Fig.~\ref{Figure_bandgap}(a)].
This characteristic length is related to the decay length
of the CBM edge-state. Figure~\ref{Figure_bandgap}(b)
shows the squared electron wavefunctions of the CBM
states of 14-aBNNR and 26-aBNNR integrated in the $yz$ plane.
These states are localized on the boron atoms near the two edges.
When the width is about 3~nm as in the 26-aBNNR,
the wavefunction from the two edges begins to decouple and thus
stabling its energy position.

In zigzag BNNRs, the boron edge and the nitrogen edge are negatively 
and positively charged (electronic plus ionic charge), respectively.
Because of this polarization,
the potential felt by electrons is higher at the boron
edge and lower at the nitrogen edge, contributing a factor which
increases the band gap of the narrow zBNNRs since
the VBM and the CBM are edge-states
localized at the nitrogen edge and the boron edge, respectively
[Fig.~\ref{Figure_z7BN}(a)].
However, as the ribbon becomes wider,
the effective polarization line charge density $\sigma_{\rm eff}$, defined as
$$\sigma_{\rm eff}\equiv\hat{x}\,\cdot\,\frac{1}{w_z h_z}\int_{{\bf r}\ \in\
{\rm unit\ cell}}d{\bf r}\ \rho({\bf r})\ {\bf r}$$
where $\rho({\bf r})$ is the total charge density including the core charge, and $h_z$ the
spatial period along the $y$ direction [see Fig.~\ref{Figure_structure}(b)],
decreases as $\sim1/w_z$ [Fig.~\ref{Figure_bandgap}(c)] due to an increased screening,
resulting in the decrease and convergence of the band gap as $w_z$ increases.

Figure~\ref{Figure_a14BN}(b) shows how the bandstructure and wavefunctions
of a 14-aBNNR change under a  0.1~eV/\AA\ transverse electric field. Owing to the Stark effect,
the wavefunctions of the highest occupied  and the lowest unoccupied states now localize
at the opposite edges where the external electrostatic potential felt by an electron
is higher and lower, respectively [right panel of Fig.~\ref{Figure_a14BN}(b)].
Thus, the band gaps of armchair BNNRs decrease when a transverse
electric field is applied [left panel of Fig.~\ref{Figure_a14BN}(b)].
A similar behavior has been predicted~\cite{khoo:2004PRB_BNNT_Stark}
and observed~\cite{ishigami:2005PRL_BNNT_Stark} in BNNTs.

Figures~\ref{Figure_z7BN}(b) and \ref{Figure_z7BN}(c)
show how the bandstructure and the edge-state
wavefunctions of a 7-zBNNR change under a $0.1$~eV/\AA\
transverse electric field.
When an electric field is applied toward $+x$ direction,
the VBM and CBM edge-state wavefunctions do not change qualitatively
[right panel of Fig.~\ref{Figure_z7BN}(b)].
Thus, for a similar reason as in the armchair BNNRs, the band gap
decreases [left panel of Fig.~\ref{Figure_z7BN}(b)].
When the electric field is applied along $-x$ direction,
the potential felt by an electron localized at the right edge (the nitrogen edge)
is decreased whereas that at the left edge (the boron edge) is increased.
Therefore, the energy gap between these two states
increases as shown in Fig.~\ref{Figure_z7BN}(c).
(Actually, the energy of the original lowest empty state has been pushed
upward by so much that the state is no longer the CBM state.)
At the same time,
the VBM states tend to delocalize
[right lower panel of Fig.~\ref{Figure_z7BN}(c)].

\begin{figure}
\includegraphics[width=1.0\columnwidth]{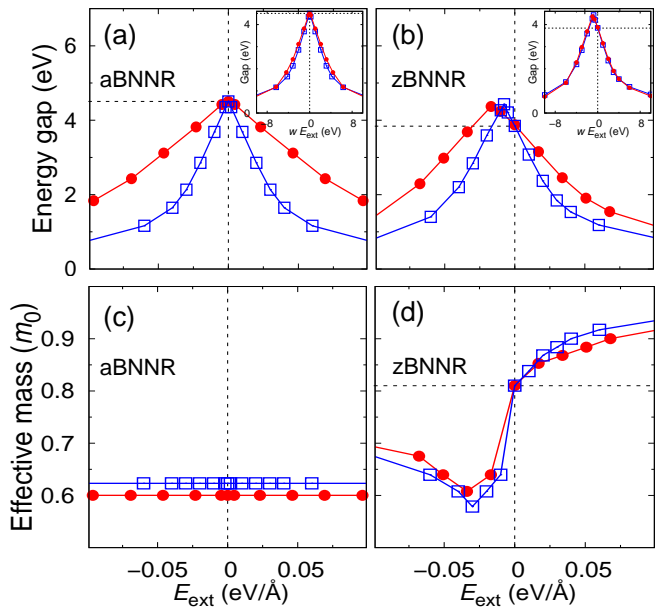}
\caption{(color online) (a) and (c): LDA energy gaps
and effective hole masses (in units of the free electron mass $m_0$)
of the highest occupied band
in 36-aBNNR (filled red circles) and 84-aBNNR
(empty blue squares) under a transverse electric field versus the field strength.
The inset in (a) shows energy gaps as a function of the
external potential difference between the two edges.
(b) and (d): Similar quantities as in (a) and (c) for 27-zBNNR
(filled red circles) and 46-zBNNR (empty blue squares), respectively.}
\label{Figure_Egap}
\end{figure}

Figures~\ref{Figure_Egap}(a) and \ref{Figure_Egap}(b) show the band gap variation of BNNRs
with field strength.
In armchair BNNRs, the band gap decreases when the field strength increases
regardless of its direction [Fig.~\ref{Figure_Egap}(a)].
A similar behavior has been observed in SW-BNNTs~\cite{khoo:2004PRB_BNNT_Stark}.
For example, for the 10~nm wide 84-aBNNR
the LDA band gap is reduced from 4.5~eV to less than 1.0~eV
under a  $0.1$~eV/\AA\ field.
In zigzag BNNRs, at {\it small} field strength, the band gap decreases when the field is along $+x$
direction but increases when the field is reversed
[Fig.~\ref{Figure_Egap}(b)]. In other words,
zigzag BNNRs show asymmetric Stark effect.
However, as the field becomes stronger, the band gap decreases
regardless of the direction.
Moreover, the band gap variations, when plotted against the difference in the external
electrostatic potential between the two edges,
fall on a universal curve for ribbons with different widths
[insets of Figs.~\ref{Figure_Egap}(a) and \ref{Figure_Egap}(b)].
This is because the gap determining
states localize on different edges as the field
becomes strong; thus, the change in their energy difference
is directly related to the potential difference
between the edges.

Figures~\ref{Figure_Egap}(c) and \ref{Figure_Egap}(d) show the effective mass
of the hole carrier at the VBM for a range of external field strength.
In armchair BNNRs, the hole mass of the VBM
is independent of the field strength.
In contrast, the corresponding effective mass in zigzag BNNRs
changes with the external field, and even more interestingly, in an asymmetric way.
In particular, when the field is along $-x$ direction,
the effective mass decreases substantially.
Within $\pm0.02$~eV/\AA\ variation of the field, the effective mass can
be varied by 50~\% from $0.6\,\,m_0$ to $0.9\,\,m_0$
where $m_0$ is the free electron mass.
In the case of electron carriers in zigzag BNNRs,
a nearly-free-electron
state~\cite{blase:1995PRB_hBN_GW,khoo:2004PRB_BNNT_Stark,park:2006PRL_BNNT_Opt} becomes
the CBM if a field stronger than a critical value, depending on the width,
is applied [see Fig.~\ref{Figure_z7BN}(c)], and
the characteristics of charge carriers change significantly.
These novel phenomena
demonstrate the possibility of tuning carrier mobilities of
doped BNNRs by applying a transverse electric field.

In summary, we have studied the electronic
properties of BNNRs as a function of width
with or without an external transverse electric fields.
The band gap of the armchair BNNR and that of the zigzag BNNR
are determined by edge-states and thus
converge to values different from that of the bulk BN sheet.
The electronic and the transport properties of BNNRs
are shown to be tunable by an external
transverse electric field. Especially, zigzag BNNRs are
shown to exhibit asymmetric response to the electric field.

{\it Additional remark}: After completion of the work and during the
preparation of this manuscript, we became aware of a related work
on similar systems from other group~\cite{zhang:2008PRB_BNNR}.

We thank A. Khoo, J. D. Sau, E. Kioupakis and F. Giustino
for fruitful discussions
and Y.-W. Son for critical reading of the manuscript.
This work was supported by NSF Grant
No. DMR07-05941 and by the Director, Office of Science, Office of Basic Energy
Sciences, Division of Materials Sciences and Engineering Division,
U.S. Department of Energy under Contract No. DE- AC02-05CH11231.
Computational resources have been provided by NPACI and NERSC.


\begin{thebibliography}{10}
\bibitem{note:bulk}
In this paper, we mean by bulk gap the electronic energy band gap of an
  infinite boron nitride sheet without edges simulated by imposing a periodic
  boundary condition.
\end{thebibliography}
\end{document}